\documentclass[aps,preprint,showpacs]{revtex4}
\usepackage{graphicx}
\usepackage{dcolumn}
\usepackage{bm}
\usepackage[latin1]{inputenc}
\usepackage{hyperref}
\usepackage{color}

\begin{document}

\preprint{}
\title{Observation of vortex formation in an oscillating trapped Bose-Einstein condensate}
\author{E. A. L. Henn$^1$}\email{ehenn@ifsc.usp.br}
\author{J. A. Seman$^1$}
\author{E. R. F. Ramos$^1$}
\author{M. Caracanhas$^1$}
\author{P. Castilho$^1$}
\author{E. P. Olímpio$^1$}
\author{G. Roati$^2$}
\author{D. V. Magalh\~{a}es$^1$}\altaffiliation[Permanent adress: ]{Escola de Engenharia de São Carlos - USP Depto. de Engenharia Mecânica
São Carlos - SP - Brazil - 13566-590}
\author{K. M. F. Magalh\~{a}es$^1$}
\author{V.S. Bagnato$^1$}
\affiliation{$^1$Instituto de F\'{\i}sica de S\~{a}o Carlos -- USP\\ C.P. 369, S\~{a}o Carlos -- SP - Brazil -- 13560-970\\ $^2$LENS and Dipartamento di Fisica, Università di Firenze,\\ INFN and CNR-INFM, 50019 Sesto Fiorentino, Italy}

\begin{abstract}

We report on the observation of vortex formation in a Bose-Einstein condensate of $^{87}Rb$ atoms. Vortices are generated by superimposing an oscillating excitation to the trapping potential introduced by an external magnetic field. For small amplitudes of the external excitation field we observe a bending of the cloud axis. Increasing the amplitude we observe formation of a growing number of vortices in the sample. Shot-to-shot variations in both vortex number and position within the condensed cloud are observed, probably due to the intrinsic vortex nucleation dynamics. We discuss the possible formation of vortices and anti-vortices in the sample as well as possible mechanisms for vortex nucleation.

\end{abstract}


\maketitle

\section{Introduction}\label{sec:intro}

Superfluidity is a remarkable signature of the quantum nature of a given physical system. This phenomenon is consequence of the macroscopic occupation of a single quantum state and reveals itself as a frictionless flow without dissipation. The superfluidity was discovered in 1938 by Kapitza \cite{kapitza} and, independently, by Allen and Misener \cite{allen} in a sample of liquid $^4He$ at a temperature of $2.17 K$. In particular, formation of quantized vortices is one of the most outstanding features of superfluids, and a considerable amount of work has been done on systems like $^4He$ or mixtures of $^4He/^3He$ \cite{Hebook}. 

The achievement of quantum degeneracy \cite{Bec} in trapped dilute atomic gases has opened new directions in the study of superfluidity and vorticity. In fact, these quantum fluids are natural testing grounds for studying properties and phenomena related to superfluidity due to the unique possibility of controlling many of the parameters of the system in an easy way. In particular, the observation of quantized vortices in a BEC \cite{firstvortex} and, more recently, in a quantum degenerate Fermi gas \cite{Fermion-Ketterle}, has been considered as an undeniable proof of their superfluid nature. 

Besides the big tunability of the different parameters of these systems, there are many advantages in realizing these studies in quantum degenerate gases. The possibility of expanding the quantum atomic fluid, together with the fact that the low density regime yields a quite large healing length which characterizes the vortex core size, make possible the direct observation of the vortex characteristic distribution, geometry and location using optical absorption techniques.

The first experimental production of a vortex in an atomic BEC was performed by Matthews \textit{et al.} \cite{firstvortex} using a phase engineering technique. In that case, rotation was dynamically induced by driving a transition between two different hyperfine spin states of the trapped atoms, using a rotating field. Vortices have also been produced by focalization of a laser beam on the condensate moving faster than the critical velocity \cite{stirring}. Stirring the condensate with a laser beam has allowed many different experiments involving vortex formation \cite{vortexformation}, determination of the transferred angular momentum to the cloud and determination of threshold-frequency during vortex excitation \cite{vortexexcitation}. Rotating the condensate with an asymmetric trapping potential \cite{assymetric potential} was the first demonstration of a purely magnetic excitation scheme quite analogous to the rotating bucket experiment with liquid He \cite{Hebook}. Also, a dynamical instability \cite{dyna} lead to the nucleation of vortices, with subsequent crystallization in a lattice configuration. Recently, vortex formation has been also observed by merging multiple trapped BECs \cite{Scherer} and by coherent transfer of orbital angular momentum from optical fields to the condensed sample \cite{LG}. Vortices in superfluids continue to generate interest both theoretically \cite{theoryonvortex} and experimentally \cite{experimentsonvortex}.

In this communication we show experimentally that vortices can be nucleated in a BEC when an oscillating spherical quadrupole field is applied to it. We are able to transfer several quanta of angular momentum to the condensate, evidenced by the appearance of many vortices in the sample. This technique is similar to the theoretical proposal by Möttönen \textit{et al.} \cite{pump} and recently by Z.F. Xu \textit{et al.}\cite{you} to pump angular momentum in a BEC using an external magnetic field to imprint local Berry phase in the sample, though we do not believe that this is the actual mechanism in our case, mainly because in both protocols an inversion of the magnetic bias field is needed and that is not the case in our experiment.

\section{Experimental Setup}\label{sec:exp}

\begin{figure} 
\centering
 \includegraphics[scale=0.55]{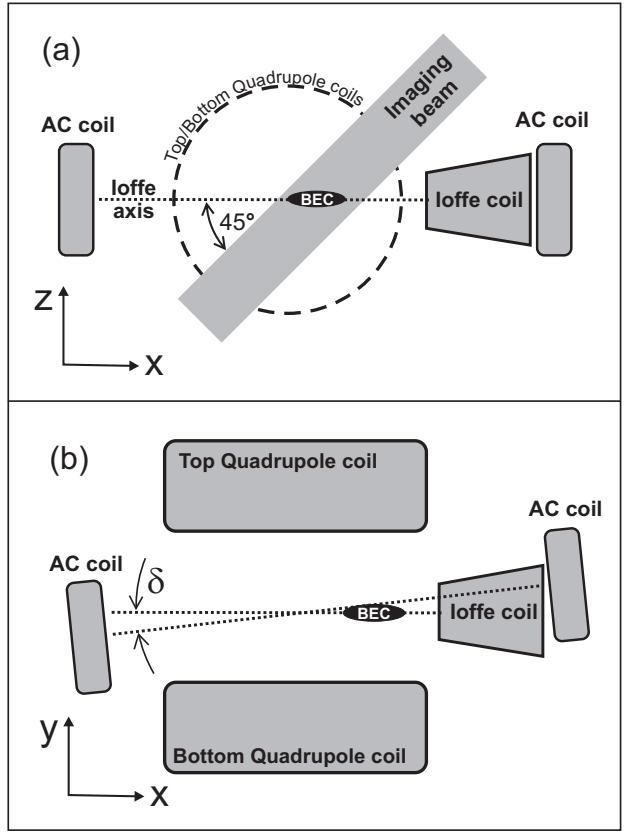}
 \caption{Schematics of (a) above view, and (b) side view of the imaging beam and the coil arrangement in our system, showing the QUIC trap coils and the ACC set of coils. The drawing is not to scale and the ACC misalignment is made bigger for clarity.}
\label{fig:oaccoil}
\end{figure}

The experimental setup to produce the quantum sample is described in greater detail in Ref.\cite{BJP}. Briefly, we collect $10^9$ $^{87}$Rb atoms at $100 \mu K$ in a MOT. The atoms are then transferred in the $\left|2,2\right\rangle$ hyperfine state and then loaded into a QUIC-type magnetic trap \cite{QUIC}. Once atoms are in the QUIC potential, RF-induced evaporation is used to obtain quantum degeneracy. The trapping potential is harmonic with frequencies given by $\omega_x=\omega_0$ and $\omega_y=\omega_z=9\omega_0$, where $\omega_0\approx2\pi\times23$Hz. We extract the parameters of the atomic cloud (number of atoms, density profile and temperature) by imaging it on a CCD camera after free fall of $15$ ms once the trapping potential is turned off. The imaging beam propagates in the $xz$ plane, making a $45^o$ angle with the Ioffe coil axis, as shown in Fig.\ref{fig:oaccoil}(a). We typically produce BEC samples containing $1-2\times10^5$ atoms.

After reaching BEC, an extra field, produced by a pair of anti-Helmholtz coils is superimposed on the QUIC trap field as shown in Fig.\ref{fig:oaccoil}(b). The axis of the extra coils is close to the trap axis (the angle between these axes is $\delta\sim 5^o$). The center of the extra field, defined by the zero-field amplitude position, is close to the QUIC trap minimum. An oscillatory current is applied to the coils, always having the same sign and always starting from zero, so we do not give an abrupt kick to the condensate in the beginning of the excitation phase ($I_{coil}=I_0\left[1-\cos(\Omega t)\right]$). The experimental sequence is then as follows: finished the evaporative cooling, the external field is turned on for tens of milliseconds. Once the external field is turned off, atoms are held in the static magnetic trap for extra 20ms prior to be released from the QUIC trap. Atoms are imaged after a 15ms time-of-flight. For this experiment we restrict ourselves to excitation periods running from zero to $60$ms and maximum amplitudes of the spatial gradient of the field up to $\sim200$ mG/cm.

\begin{figure} 
\centering
 \includegraphics[scale=0.35]{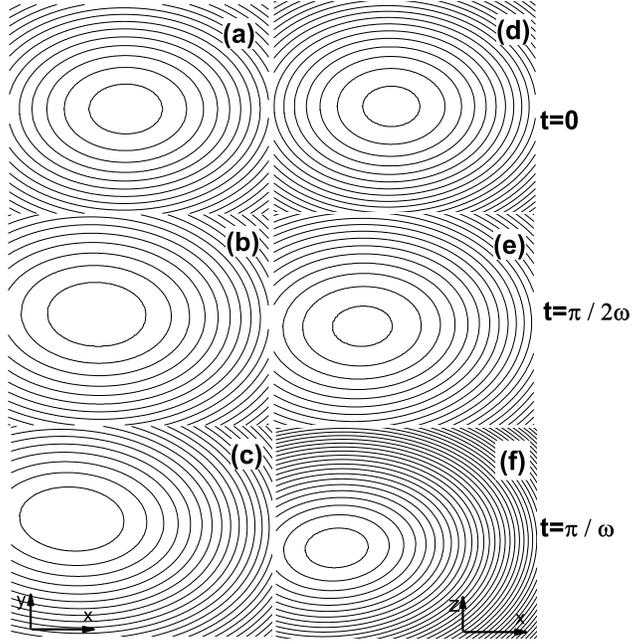}
 \caption{Simulated movement/deformation of the trapping potential under the influence of the external oscillatory field. (a), (b), (c) are the XY plane and (d), (e), (f) represent the XZ plane. Images (a) and (d) are at t=0, the starting point of the excitation, (b) and (e) are at midway between zero and the maximum amplitude of the external field ($t=\pi/2\omega$) and (c) and (f) are at maximum amplitude field ($t=\pi/ \omega$). The field of view is the same for all images.}
\label{simul}
\end{figure}

We indicate the set of excitation coils as AC-coils (ACC). The effect of field added by the ACC over the static trap is to produce a combination of \textit{translation}, \textit{i. e.} a displacement of the minimum trap, \textit{rotation}, due to the angular dislocation of the coils, and a \textit{shape deformation}, due to the asymmetry caused by the combination of the ACC field with the QUIC field. A simulation of the field for experimental parameters close to the real ones is shown in Fig.\ref{simul}. 

\section{Vortex Observation}\label{sec:vortex}

The spatial gradient of the field produced by the ACC is small, on the order of tens to hundreds of miliGauss/cm for the axial gradient, which in amplitude is comparable with the field variation experienced by the atoms in the BEC. We will show that for certain values of amplitude and frequency, this configuration transfers angular momentum to the BEC and hence produces vortices. 

\begin{figure} 
\centering
 \includegraphics[scale=0.4]{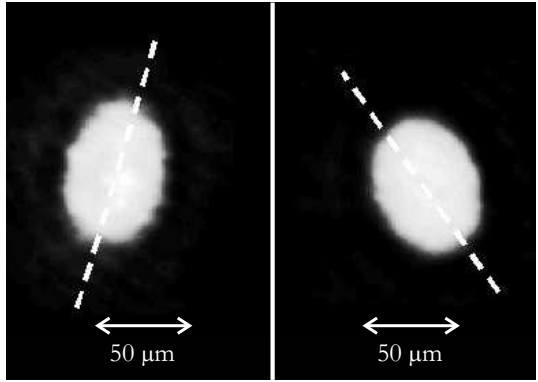}
 \caption{Observation of cloud axis bending oscillations after low amplitude ($<40$ mG/cm) application of the ACC field. Each image is the result of a different run of the experiment under similar experimental conditions and shows the optical density of the sample after 15ms of free expansion}
\label{fig:bending}
\end{figure}

A small amplitude of excitation ($<40$ mG/cm) produces only a bending oscillation of the trapped cloud axis as shown in Fig.\ref{fig:bending}. This tilting on the superfluid axis shows that excitation by the oscillatory field is able to mechanically transfer angular momentum to the atomic cloud. This effect can be related to scissor modes extensively investigated in references \cite{scissor, scissor2}.

\begin{figure} 
\centering
 \includegraphics[scale=0.5]{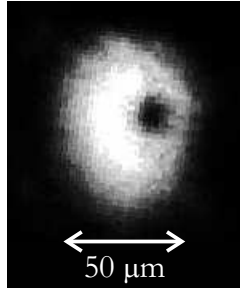}
 \caption{Optical density of the sample showing a typical single vortex pattern after application of the ACC oscillatory field after 15ms of free expansion.}
\label{fig:onevortex}
\end{figure}

\begin{figure}  
\centering
 \includegraphics[scale=0.4]{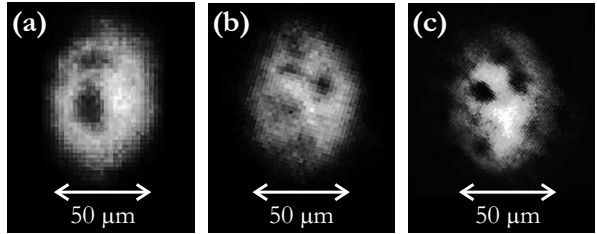}
 \caption{Typical atomic optical density images showing the appearance of (a) two and (b), (c) multiple vortices patterns after different excitation conditions. Images were taken after 15ms of free expansion.}
\label{fig:vortexfamily}
\end{figure}

The oscillatory excitation nucleates vortices only after reaching certain amplitude for the ACC field. For a fixed time of excitation (20ms), $\Omega=2\pi\times200$Hz, and amplitudes up to 40 mG/cm of the axial field of the ACC coils, we observe no evidence of vortex formation. Nevertheless, the stirring of the whole cloud as in Fig.\ref{fig:bending} is present. Increasing the amplitude further, the appearance of a single vortex type pattern starts to take place, as shown in Fig.\ref{fig:onevortex}. Up to about 90 mG/cm we observe either zero or one vortex when the same conditions are considered. The successive repetition of the experiment under the interval between 40 and 90 mG/cm reveals an average number of vortex formation of $0.6\pm 0.3$. We call this as a single vortex zone of amplitude. For amplitudes larger than 90 mG/cm and up to 160 mG/cm, most of the images show the presence of two or more vortex-type profiles, with a predominance of three vortices and occasional formation of one or more than three vortices. Typical images acquired in this regime are shown in Fig.\ref{fig:vortexfamily}. Again, several repetitions of the experiment within this amplitude range under equal experimental conditions allows to observe a multiple vortex zone, with an average number of vortices of $2.6\pm 1.2$. A considerable increase in the number of vortices is observed with larger oscillation amplitudes. We will not address this region in this communication since the vortices patterns observed are not regular nor the cloud seems to behave in the same way. These evidences might indicate that we access other regimes of vortex formation and equilibrium, not yet understood and out of the scope of this work.

The variation of vortex number observed in our system when operating at equal conditions is not clear. It could be related to the mechanisms involved in the vortex formation and eventual decay during our excitation or it can be due to some out of control instability generated by the presence of the ACC field and its mutual inductance with the QUIC trap coils. Another important feature that has to be taken into account for explaining this fluctuation is that there can be a kind of energy degeneracy among different vortex states, resulting in equal probability of formation at equivalent experimental conditions.

\begin{figure}
\centering
 \includegraphics[scale=0.35]{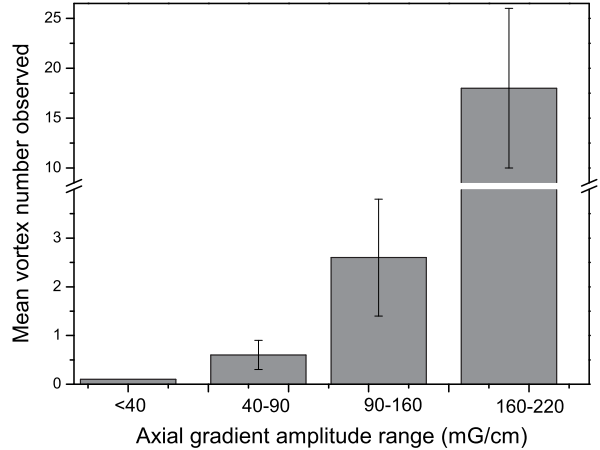}
 \caption{Mean number of vortex observed in condensed clouds as a function of excitation amplitude range for a fixed excitation period of 20ms. After excitation atoms are left trapped for additional 20ms.}
\label{fig:vortexhistogram}
\end{figure}

Fig.\ref{fig:vortexhistogram} summarizes our experimental observations and clearly shows that we increase the angular momentum transfer to the condensate as the amplitude of the excitation is increased. This is evidenced by a fast growing average number of vortices as the amplitude of the ACC field increases.

Our experimental data are quite similar to the patterns observed on the seminal observations of vortices in Bose-condensate gases \cite{firstvortex, vortexformation}. The clear difference from our work and previously related work is that our vortices observed after TOF do not seem to be symmetrically distributed along the condensate \cite{vortexexcitation}. Rather, from shot-to-shot they appear at different relative spatial positions. In the case of a single vortex, we have observed its position at the center of the condensate or towards the edges of it. A possible reason for these unusual patterns observed is that we wait only 20ms with the atoms held in the magnetic trap prior to release and the well-known regular structures typically need more time (on the order of hundreds of milliseconds) to be formed. In other hand, our results resemble the results observed in Ref.\cite{Scherer}. Possibly, the similarities can be regarded to the stochastic nature of the excitation. In their case, it is due to the relative phase of the merging condensates. In our case, two possibilities arise. First, it is possibly due to the fact that one cannot say in which direction of the oscillatory motion of the condensate the vortex is excited and it is likely to vary from shot-to-shot. Second, if we are really imprinting Berry phase on the condensate, the decay mechanisms might be responsible for these observations and unusual spatial distributions. Nevertheless, to imprint Berry phase on the BEC, the protocols described in \cite{pump, you} propose a continuous exchange from quadrupolar to hexapolar potentials so breaking the time-reversal symmetry of the processes of winding and unwinding the condensate. That procedure gives the sample net angular momentum and promotes the appearance of vortices. In addition, in both vortex pump protocols, the inversion of the magnetic bias field is an essential requirement. The non-desired effects due to that change like spin-flips and subsequent loss of atoms from the trap are overcome by adding an extra optical potential. In our case, there are no such changes. The potential remains harmonic, without inversion of the bias field as it is shown in the simulation of Fig.\ref{simul}. Increasing the time between the end of excitation and releasing the atoms from the trap would help to discriminate the dynamics of the vortex patterns observed. In particular, the observation of heating might indicate the decay mechanism of the vortices observed. For this set of measurements, we did not investigate for much longer times of evolution of the BEC, but for slightly longer times ($\approx$30 ms) no considerable heating is observed. 

In order to try to discriminate which is the possible mechanism of vortex nucleation in our experiment we performed 2D numerical simulations. The preliminary simulations consisted in considering the time dependent Gross-Pitaevskii equation (GPE) plus an external potential similar to the one generated by our coils. We find a rather complex behavior of the condensate dynamics compared to the cases studied so far of vortex nucleation in the presence of a weak elliptical deformation of a rotating trap, where vortices emerge in a rather straightforward way \cite{tsubota, gardiner}. This is due to the elongated geometry of our condensate, that is subjected to large deformations for our experimental procedure, making the numerical analysis of possible formation of vortices or turbulent regimes \cite{tsubota2} more difficult. In our case, the effect of the thermal cloud is included in an effective way by means of noise on the initial wavefunction. However, we are not able to simulate explicitly the friction between the condensed and thermal clouds. In fact, we believe that a possible mechanism of formation of the vortices might be related to the relative movement of the condensed and thermal components subjected to the external field. This phenomenon is well-known in the scope of fluid interface theory and experiments and is named Kelvin-Helmholtz instabilities \cite{KH}. These instabilities occur in the surface of two fluids that have a relative velocity and give rise to the formation of vortices in the interface of them. This phenomenon has been studied in mixtures of superfluid-normal fluid liquid Helium \cite{KH superfluid} and has been theoretically investigated in this context \cite{KH theory}. Nevertheless, the Kelvin-Helmholtz instabilities have never been directly observed as interface excitations in quantum atomic fluids, though evidence of Kelvin modes of a vortex line have been observed \cite{kelvinmodes} in the decay of a counter rotating mode in the presence of a quantized vortex in a Bose-Einstein condensate. 

\begin{figure}
\centering
 \includegraphics[scale=0.8]{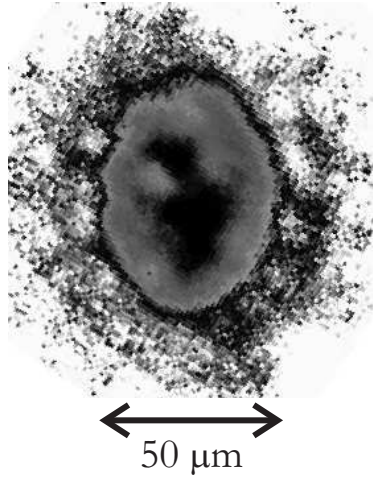}
 \caption{Atomic optical density image of the atomic cloud from Fig.\ref{fig:vortexfamily}(c) with a different contrast, showing round structures around the condensed component.}
\label{fig:border}
\end{figure}

We claim that this is a possible mechanism of formation of our vortices because, if one changes the contrast of our images to highlight the interface between the condensed component to the normal component of the gas, round structures can be seen all around the condensed cloud, as shown in Fig.\ref{fig:border}. This image is the one shown in Fig.\ref{fig:vortexfamily}(c) where only the contrast has been modified. This experimental observation performed just after excitation reveals large population of vortex-like excitations at the interface and certainly indicates towards the mechanism proposed.

Currently, more thorough simulations are under way to confirm or deny these hypotheses as well as trying to apply the standard Kelvin-Helmholtz instabilities theory to quantum atomic fluids in order to justify and clarify the true nature of the vortex nucleation observed in our sample. The full theoretical treatment is out of the scope of this communication and will be addressed in a future publication.

An important peculiarity of our experiment is that our way to produce vortices is quite different from the conventional introduction of rotation by a ``spoon'', which always results in a single sense of rotation and therefore the formation of vortices with the same circulation sign. For our case, the oscillation in the excitation represents equal chances to form vortex (positive-circulation) and anti-vortex (negative circulation). In fact, we believe that both types are being simultaneously produced in the experiment. This fact may well be part of the reasons for observations of such large fluctuations in the response of the system under equivalent experimental conditions.

\section{Conclusions}\label{sec:conclusions}

To summarize, we produced vortices in a BEC of $^{87}Rb$ atoms by superimposing an oscillatory field to the magnetic trap. By changing the amplitude of the external excitation we can observe a regime without vortices and another regime where several vortices are nucleated. The mechanism responsible for the vortex nucleation due to oscillations seems to be related with the presence of a thermal component and its relative motion with the condensed fraction. At present, we are performing numerical simulations for proving this hypothesis.

Another important characteristic of our experiment is that the oscillatory motion conditions can generate both circulation sign vortices and therefore creating interesting conditions for vortex/anti-vortex experiments. A systematic investigation increasing the in-trap evolution time of the sample should reveal details on the relaxation processes, specially the possible vortex-antivortex annihilation that would result in heating or the appearance of more regular patterns.

Finally, as previously shown, increasing the amplitude of the excitation, the number of vortices nucleated increases dramatically, opening up the possibility of generating turbulent regimes in the quantum fluid. Quantum turbulence was discovered in superfluid $^4He$ more than 50 years ago \cite{turbulence}. However, it has never been observed in a quantum degenerate gas. A turbulent regime in such a system would represent a new scenario for studying this phenomenon, bringing new insights on it.  

\acknowledgments

We acknowledge J. Dalibard (ENS-France), S.R. Muniz (NIST) and V.I. Yukalov (Dubna-Russia) for fruitful discussions. We thank M. Modugno for the preliminary 2D simulation and useful discussions on the theoretical investigation of this subject. This work was supported by FAPESP (program CEPID), CAPES and CNPq.


\begin{thebibliography}{}






\bibitem{kapitza} P. L. Kapitza, Nature \textbf{141} (1938) 913.

\bibitem{allen} J. F. Allen, A. D. Misener, Nature \textbf{141} (1938) 75.

\bibitem{Hebook} R. J. Donelly, Quantized Vortices in Helium II, Cambridge University Press, Cambridge, 1991.

\bibitem{Bec} M. H. Anderson, J. R. Ensher, M. R. Matthews, C. E. Wieman, E. A. Cornell, Science \textbf{269} 198-201 (1995).

\bibitem{firstvortex} M. R. Matthews, B. P. Anderson, P. C. Haljan, D. S. Hall, C. E. Wieman, E. A. Cornell, Phys. Rev. Lett. \textbf{83}, 2498-2501 (1999).

\bibitem{Fermion-Ketterle} M. W. Zwierlein, J. R. Abo-Shaeer, A. Schirotzek, C. H. Schunck, W. Ketterle, Nature \textbf{435} 1047-1051 (2005).

\bibitem{stirring} S. Inouye, S. Gupta, T. Rosenband, A. P. Chikkatur, A. G\"orlitz, T. L. Gustavson, A. E. Leanhardt, D. E. Pritchard, W. Ketterle, Phys. Rev. Lett. \textbf{87}, 080402 (2001).

\bibitem{vortexformation} K. W. Madison, F. Chevy, W. Wohlleben, J. Dalibard,  Phys. Rev. Lett. \textbf{84} 806-809 (2000).

\bibitem{vortexexcitation} F. Chevy, K. W. Madison, J. Dalibard, Phys. Rev. Lett. \textbf{85} 2223-2227 (2000). 

\bibitem{assymetric potential} E. Hodby, G. Hechenblaikner, S. A. Hopkins, O. M. Marag\`o, C. J. Foot, Phys. Rev. Lett. \textbf{88} 010405 (2001).

\bibitem{dyna} K. W. Madison, F. Chevy, V. Bretin, J. Dalibard,  Phys. Rev. Lett. \textbf{86} 4443(2001); S. Sinha and Y. Vastin, Phys. Rev. Lett. \textbf{87} 190402 (2001)

\bibitem{Scherer}  D. R. Scherer, C. N. Weiler, T. W. Neely, B. P. Anderson, Phys. Rev. Lett. \textbf{98} 110402 (2007).

\bibitem{LG} M. F. Andersen, C. Ryu, Pierre Clad\'e, Vasant Natarajan, A. Vaziri, K. Helmerson, W. D. Phillips, Phys. Rev. Lett. \textbf{97}  (2006) 170406; K. C. Wright, L. S. Leslie, N. P. Bigelow, Phys. Rev. A \textbf{77} 041601(R) (2008).

\bibitem{theoryonvortex} A. Fetter, arXiv:0801.2952v1 (2008).

\bibitem{experimentsonvortex}	K. Kasamatsu, M. Tsubota, Prog. Low Temp. Phys. \textbf{16}, 349-401 (2008).

\bibitem{pump}  M. M\"ott\"onen, V. Pietil\"a, S. M. M. Virtanen, Phys. Rev. Lett. \textbf{99} 250406 (2007). 

\bibitem{you} Z.F. Xu, P. Zhang, C. Ramam and L. You, Phys. Rev. A \textbf{78}, 043606 (2008)

\bibitem{BJP} E. A. L. Henn, J. A. Seman, G. B. Seco, E. P. Olimpio, P. Castilho, G. Roati, D. V. Magalh\~aes, K. M. F. Magalh\~aes, V. S. Bagnato, Braz. Journ. Phys. \textbf{38} 279-286 (2008).


\bibitem{QUIC} T. Esslinger, I. Bloch, T. W. H\"ansch, Phys. Rev. A \textbf{58} R2664-R2667 (1998).


\bibitem{scissor} O. M. Marag\`o, S. A. Hopkins, J. Arlt, E. Hodby, G. Hechenblaikner, C. J. Foot, Phys. Rev. Lett. \textbf{84} 2056-2059 (2000).

\bibitem{scissor2} M. Modugno, G. Modugno, G. Roati, C. Fort, M. Inguscio, Phys. Rev. A \textbf{67} 023608 (2003).





\bibitem{tsubota} K. Kasamatsu, M. Tsubota, M. Ueda, Phys. Rev. A \textbf{67} 033610 (2003); K. Kasamatsu, M. Machida, N. Sasa, M. Tsubota, Phys. Rev. A \textbf{71} 063616 (2005).

\bibitem{gardiner} T. M. Wright, R. J. Ballagh, A. S. Bradley, P. B. Blakie, C. W. Gardiner, Phys. Rev. A \textbf{78}, 063601 (2008).

\bibitem{tsubota2} M. Kobayashi, M. Tsubota, Phys. Rev. A \textbf{76} 045603 (2007).

\bibitem{KH} G. E. Volovik, JETP Letters \textbf{75} 418-422 (2002).

\bibitem{KH superfluid} R. Blaauwgeers, V. B. Eltsov, G. Eska, A. P. Finne, R. P. Haley, M. Krusius, J. J. Ruohio, L. Skrbek, G. E. Volovik, Phys. Rev. Lett. \textbf{89} 155301 (2002).

\bibitem{KH theory} S. E. Korshunov, JETP Lett. \textbf{75},423 (2002).

\bibitem{kelvinmodes} V. Bretin, P. Rosenbusch, F. Chevy, G.V. Shlyapnikov and J. Dalibard, Phys. Rev. Lett., \textbf{90}, 100403 (2003)

\bibitem{turbulence} W. F. Vinen, Proc. R. Soc. A \textbf{240} 114-127 (1957); W. F. Vinen, Proc. R. Soc. A \textbf{240} 128-143 (1957); W. F. Vinen, Proc. R. Soc. A \textbf{242} 493-515 (1957).

\end{thebibliography}
\end{document}